\documentstyle[12pt,a4]{article}
\title{Relativistic mean-field description of light $\Lambda$ hypernuclei
with large neutron excess}

\author{D. Vretenar, W. P\"oschl,
G.A. Lalazissis and P. Ring\\
Physik-Department, Technische Universit\"at  M\"unchen,\\
D-85748 Garching, Germany}
\begin{document}
\maketitle
\begin{abstract}
The Relativistic Hartree Bogoliubov
model in coordinate space, with finite range pairing 
interaction, is applied to the description 
of $\Lambda$-hypernuclei with a large neutron excess.
The addition of the $\Lambda$ hyperon to Ne isotopes
with neutron halo can shift the neutron drip by 
stabilizing an otherwise unbound core nucleus.
The additional binding of the halo
neutrons to the core originates
from the increase in magnitude of the spin-orbit term.
Although the $\Lambda$  produces only a 
fractional change in the central mean-field potential,
through a purely relativistic effect it increases
the spin-orbit term which binds the outermost neutrons.
\end{abstract}


\vskip 1.5cm 
\noindent 

The production mechanisms, spectroscopy, and decay modes
of hypernuclear states have been the subject of many
theoretical studies. Extensive reviews of 
the experimental and theoretical status of strange-
particle nuclear physics can be found in 
Refs.~\cite{CD.89,DMG.89,BMZ.90}.
The most studied hypernuclear system consists of a single
$\Lambda$ particle coupled to the nuclear core. And 
although strangeness in principle can be used as a 
measure of quark deconfinement in nuclear matter, a 
single $\Lambda$ behaves essentially as a distinguishable
particle in the nucleus. Theoretical models used in 
studies of hypernuclei extend from nonrelativistic 
approaches based on OBE models for $\Lambda$-N 
interaction, to the relativistic mean field approximation
and quark-meson coupling models. However, our knowledge 
of the $\Lambda$-nucleus interaction, and of 
hypernuclear systems in general, is restricted to the 
valley of $\beta$-stability.

In recent years the study 
of the structure of exotic nuclei, produced by 
radioactive nuclear beams, has become one of the
most active fields in nuclear physics. The structure 
of nuclei far from the stability line presents many
interesting phenomena. In particular, in the present 
work we consider the 
extremely weak binding of the outermost 
nucleons, large spatial dimensions and the
coupling between bound states and the particle continuum.
By adding either more protons or neutrons, the particle 
drip lines are reached. Nuclei beyond the drip lines 
are unbound with respect to nuclear emission. Exotic 
nuclei on the neutron-rich side are especially important
in nuclear astrophysics. They are expected to play an 
important role in nucleosynthesis by neutron capture 
(r-processes). Knowledge of their structure and
properties would help the determination of astrophysical
conditions for the formation of neutron-rich stable
isotopes. On the neutron-rich side, the drip line has 
only been reached for very light nuclei. 

It has been recently suggested~\cite{Maj.95} that a study 
of $\Lambda$-hypernuclei with a large neutron excess
could also display interesting phenomena. On one  
hand such hypernuclei, corresponding to core nuclei 
which are unbound or weakly bound, are of considerable
theoretical interest. Some possibilities for unusual
light hypernuclei were already analyzed in the early 
work of Dalitz and Levi Setti~\cite{Dal.63}. On the other
hand, one could speculate on the possible role of 
neutron-rich $\Lambda$-hypernuclei in the process of 
nucleosynthesis. The $\Lambda$ particle provides the 
nuclear core with additional binding. Even-core nuclei 
that are either weakly bound or unbound attain normal
binding~\cite{Gal.75}. 

In Refs.~\cite{PVL.97,LVR.97} we 
have investigated, in the framework of relativistic mean-
field theory, light nuclear systems with large neutron 
excess. For such nuclei the separation energy of the
last neutrons can become extremely small. The Fermi 
level is found close to the particle continuum, and
the lowest particle-hole or particle-particle modes
couple to the continuum. In Ref.~\cite{PVL.97}   
the Relativistic Hartree Bogoliubov (RHB) model 
has been applied, in the 
self-consistent mean-field approximation, to the 
description of the neutron halo in the mass region above 
the s-d shell. As an extension of non-relativistic
HFB-theory~\cite{Doba.84}, the RHB theory in coordinate 
space provides a unified description of mean-field 
and pairing correlations. Pairing correlations and the 
coupling to particle continuum states have been  
described by finite range two-body Gogny-type interaction.
Finite element methods have been used 
in the coordinate space discretization of the coupled 
system of Dirac-Hartree-Bogoliubov 
and Klein-Gordon equations. Solutions in coordinate space
are essential for the correct description of the coupling 
between bound and continuum states. Calculations have 
been performed for the isotopic chains of Ne and C 
nuclei. Using the NL3~\cite{LKR.96} parameter set for 
the mean-field Lagrangian, and D1S~\cite{BGG.84} 
parameters for the Gogny interaction, we have found
evidence for the occurrence of multi-neutron halo in 
heavier Ne isotopes. We have shown that the 
properties of the 1f-2p orbitals near the Fermi level and 
the neutron pairing interaction play a crucial role in 
the formation of the halo. In the present work we 
essentially repeat the calculations of Ref.~\cite{PVL.97}
for the Ne isotopes, but we add a $\Lambda$ particle 
to the system. We are interested in the effects that
the $\Lambda$ particle in its ground state has on the 
core halo-nucleus. 

Relativistic mean-field models have been successfully applied in
calculations of nuclear matter and properties of finite nuclei throughout
the periodic table~\cite{Rin.96}.
The model describes the nucleus as
a system of Dirac nucleons which interact in a relativistic covariant manner
the isoscalar scalar $\sigma$-meson,
the isoscalar vector $\omega$-meson and the isovector vector $\rho$-meson.
The photon field ~$(A)$ accounts for the electromagnetic interaction. 
For hypernuclear systems the original model has to be extended
to the strange particle sector. In particular, the effective Lagrangian for
$\Lambda$ hypernuclei reads
\begin{eqnarray}
\label{equ.1}
{\cal L}&=&\bar\psi\left(\gamma(i\partial-g_\omega\omega
-g_\rho\vec\rho\vec\tau-eA)-m-g_\sigma\sigma\right)\psi
\nonumber\\
&&+\frac{1}{2}(\partial\sigma)^2-U(\sigma )
-\frac{1}{4}\Omega_{\mu\nu}\Omega^{\mu\nu}
+\frac{1}{2}m^2_\omega\omega^2\nonumber\\
&&-\frac{1}{4}{\vec{\rm R}}_{\mu\nu}{\vec{\rm R}}^{\mu\nu}
+\frac{1}{2}m^2_\rho\vec\rho^{\,2}
-\frac{1}{4}{\rm F}_{\mu\nu}{\rm F}^{\mu\nu}
\nonumber\\
&&+\bar\psi_{\Lambda}\left(\gamma(i\partial-g_{\omega\Lambda}\omega)
- m_\Lambda - g_{\sigma\Lambda}\sigma \right)\psi_\Lambda.
\end{eqnarray}

The Dirac spinors $(\psi )$ and $\psi_\Lambda$ denote the nucleon and 
the $\Lambda$ particle, respectively.  
Coupling constants $g_\sigma$, $g_\omega$, $g_\rho$,
and unknown meson masses
$m_\sigma$, $m_\omega$, $m_\rho$ are parameters,
adjusted to fit data on nuclear matter and finite nuclei.
The model includes the nonlinear self-coupling of the
$\sigma$-field (coupling constants $g_2$, $g_3$)
\begin{equation}
U(\sigma)~=~\frac{1}{2}m^2_\sigma\sigma^2+\frac{1}{3}g_2\sigma^3+
\frac{1}{4}g_3\sigma^4.
\end{equation}
Since the $\Lambda$ particle is neutral and isoscalar, 
it only couples to the $\sigma$ and $\omega$ mesons  
(coupling constants $g_{\sigma\Lambda}$ and $g_{\omega\Lambda}$).
We only consider the $\Lambda$ in the $1s$ state, and therefore
do not include the tensor $\Lambda$ - $\omega$ interaction..
The generalized single-particle hamiltonian of HFB theory
contains two average potentials: the self-consistent field 
$\hat\Gamma$ which encloses all the long range {\it ph} correlations,
and a pairing field $\hat\Delta$ which sums up the 
{\it pp}-correlations. 
In the Hartree approximation for the self-consistent 
mean field, the Relativistic Hartree-Bogoliubov (RHB) equations read
\begin{eqnarray}
\left( \matrix{ \hat h_D -m- \lambda & \hat\Delta \cr
                -\hat\Delta^* & -\hat h_D + m +\lambda
                 } \right) \left( \matrix{ U_k \cr V_k } \right) =
E_k\left( \matrix{ U_k \cr V_k } \right).
\end{eqnarray}
where $\hat h_D$ is the single-nucleon Dirac hamiltonian, 
and $m$ is the nucleon mass.
$U_k$ and $V_k$ are  quasi-particle Dirac spinors, and $E_k$ denote 
the quasi-particle energies.
The Dirac equation for the $\Lambda$ particle
\begin{equation}
\label{equ.5}
\bigl[-i{\bf\alpha\nabla}+\beta (m_\Lambda+g_{\sigma\Lambda}\sigma({\bf r})) +
g_{\omega\Lambda}\omega^0({\bf r})\bigr]\psi_\Lambda
= \epsilon_\Lambda\psi_\Lambda
\end{equation}

The RHB equations for the nucleons and the Dirac equation 
for the $\Lambda$ are solved self-consistently,
with potentials determined in the mean-field approximation from 
solutions of Klein-Gordon equations for mesons
and Coulomb field:
\begin{eqnarray}
\bigl[-\Delta + m_{\sigma}^2\bigr]\,\sigma({\bf r})&=&
-g_{\sigma}\,
\sum\limits_{E_k > 0} V_k^{\dagger}({\bf r})\gamma^0 V_k({\bf r})
-g_2\,\sigma^2({\bf r})-g_3\,\sigma^3({\bf r})
\nonumber \\
&\,&-g_{\sigma\Lambda}\psi_\Lambda^{\dagger}({\bf r})\gamma^0
\psi_\Lambda({\bf r})\, \\
\bigl[-\Delta + m_{\omega}^2\bigr]\,\omega^0({\bf r})&=&
\sum\limits_{E_k > 0} V_k^{\dagger}({\bf r}) V_k({\bf r})
+g_{\omega\Lambda}\psi_\Lambda^{\dagger}({\bf r})
\psi_\Lambda({\bf r})\, \\
\bigl[-\Delta + m_{\rho}^2\bigr]\,\rho^0({\bf r})&=&
\sum\limits_{E_k > 0} V_k^{\dagger}({\bf r})\tau_3 V_k({\bf r}), \\
-\Delta \, A^0({\bf r})&=&
\,\sum\limits_{E_k > 0} V_k^{\dagger}({\bf r}) {{1-\tau_3}\over 2} 
V_k({\bf r}).
\end{eqnarray}
The sums run over all positive energy states.
The system of equations
is solved self-consistently in coordinate space
by discretization on the finite element mesh. 
In the coordinate space representation of the pairing field 
$\hat\Delta $, the kernel of the integral operator is
\begin{equation}
\Delta_{ab} ({\bf r}, {\bf r}') = {1\over 2}\sum\limits_{c,d}
V_{abcd}({\bf r},{\bf r}') {\bf\kappa}_{cd}({\bf r},{\bf r}').
\end{equation}
where 
$V_{abcd}({\bf r},{\bf r}')$ are matrix elements of a general 
two-body pairing interaction and 
${\bf\kappa}_{cd}({\bf r},{\bf r}')$, is 
the pairing tensor, defined as
\begin{equation}
\label{equ.4}
{\bf\kappa}_{cd}({\bf r},{\bf r}') := 
\sum_{E_k>0} U_{ck}^*({\bf r})V_{dk}({\bf r}').
\end{equation}
The integral operator $\hat\Delta$ acts on the wave function
$V_k({\bf r})$:
\begin{equation}
\label{equ.2.4}
(\hat\Delta V_k)({\bf r}) 
= \sum_b \int d^3r' \Delta_{ab} ({\bf r},{\bf r}') V_{bk}({\bf r}'). 
\end{equation}
In the particle-particle ($pp$) channel
the pairing interaction is approximated by the finite
range two-body Gogny interaction
\begin{equation}
V^{pp}(1,2)~=~\sum_{i=1,2}
e^{-[ ({\bf r}_1- {\bf r}_2) 
 / {\mu_i} ]^2}\,
(W_i~+~B_i P^\sigma 
-H_i P^\tau -
M_i P^\sigma P^\tau),
\end{equation}
with the parameters 
$\mu_i$, $W_i$, $B_i$, $H_i$ and $M_i$ $(i=1,2)$.

As in Ref.~\cite{PVL.97}, the even-even Ne isotopes
have been calculated with the NL3~\cite{LKR.96} effective 
interaction for the mean-field Lagrangian, 
and the parameter set D1S~\cite{BGG.84} has been used for
the finite range pairing force. The coupling constants for
the $\Lambda$ particle are from Ref.~\cite{MJ.94}, where
the relativistic mean-field theory
was used to study  
characteristics of $\Lambda$, $\Sigma$ and $\Xi$ hypernuclei.
While the values for the $g_{\omega Y}$ coupling constants 
were determined from the naive quark model, that is 
$g_{\omega \Lambda} = \frac{2}{3} g_{\omega N}$;
the values of $g_{\sigma Y}$ were deduced from the 
available experimental information of hyperon binding
in the nuclear medium. For the $\Lambda$ hyperon 
$g_{\sigma \Lambda}$ was 
fitted to reproduce the 
binding energy of a $\Lambda$ in the $1s$ state 
of $^{17}_{\Lambda}$O: $g_{\sigma \Lambda} =
0.621 g_{\sigma N}$. The coupling constant 
determined from only this experimental quantity
gives a reasonable description of binding energies 
in $\Lambda$ hypernuclei for a wide range of mass number.

In Ref.~\cite{PVL.97} we have shown that the neutron
$rms$ radii of the Ne isotopes follow the mean-field 
N$^{1/3}$ curve up to N $\approx$ 22. 
For larger values of N both neutron and matter $rms$ radii
display a sharp increase, while the proton radii stay 
practically constant. The sudden increase in neutron 
$rms$ radii has been interpreted as evidence for the 
formation of a multi-particle halo. The phenomenon 
was also clearly seen in the plot of 
proton and neutron density distributions. 
The proton density profiles do not change with the number
of neutrons, while the neutron density distributions display
an abrupt change between $^{30}$Ne and $^{32}$Ne.
The microscopic origin of the neutron halo has been found in
a delicate balance of the self-consistent mean-field 
and the pairing field. This is shown in Fig. 1a, where
we display the neutron single-particle states 
1f$_{7/2}$, 2p$_{3/2}$ and  2p$_{1/2}$ in the canonical 
basis, and the Fermi energy as functions of the mass number A. 
For A$\leq$32 (N$\leq$ 22) the triplet of states is high in the continuum,
and the Fermi level uniformly increases toward zero. The triplet 
approaches zero energy, and a gap is formed between these
states and all other states in the continuum. The shell
structure dramatically changes at N$\geq$ 22. Between
A=32 (N = 22) and A=42 (N = 32) the Fermi level is practically constant
and very close to the continuum. The
addition of neutrons in this region of the drip
does not increase the binding. Only the
spatial extension of neutron distribution displays an increase.
The formation of the neutron halo is related to the
quasi-degeneracy of the triplet of states 1f$_{7/2}$, 2p$_{3/2}$
and  2p$_{1/2}$. The pairing interaction promotes neutrons from the
1f$_{7/2}$ orbital to the 2p levels. Since these levels are
so close in energy, the total binding energy does not change
significantly. Due to their small centrifugal barrier,
the 2p$_{3/2}$ and 2p$_{1/2}$ orbitals form the halo. The last bound
isotope is $^{40}$Ne. For N $\geq$ 32 (A=42) the neutron Fermi level
becomes positive, and heavier isotopes are not bound any
more. 

In the present work we have repeated the calculation of 
Ne isotopes, but a $\Lambda$ hyperon has been added to 
the even-even cores. As one would expect, there are
no excessive changes in the bulk properties.
For example, the neutron $rms$ radii are reduced on the 
average by 2\%. Therefore we do not display changes 
in macroscopic quantities, but going to the microscopic
level, in Fig. 1b we illustrate the effect of the 
$\Lambda$ hyperon on the triplet of neutron states 
that form the halo: 1f$_{7/2}$, 2p$_{3/2}$
and  2p$_{1/2}$, and on the Fermi level. The energies 
are displayed as function of the core mass number
A$_{c}$. Due to the extra binding provided by the 
$\Lambda$, the single-neutron energies and the 
Fermi level are lower. The most important effect
that we observe, however, is that the Fermi level is 
negative for the isotope $^{42+\Lambda}$Ne. Without 
the $\Lambda$, the nucleus $^{42}$Ne was unbound. 
The presence of the strange baryon stabilizes the 
otherwise unbound core. This could have interesting 
consequences for the process of nucleosynthesis, 
especially r-processes.
It should be noted, however, that heavier nuclei, and in 
particular those with atomic number around Z=40, are 
important in the rapid neutron-capture mechanisms. 
In a very recent study~\cite{MR.97}, it has been shown 
that RHB theory predicts the last bound Zr isotope
to be $^{138}$Zr (N=98). Therefore we have calculated 
the ground state of the hypernucleus
$^{140+\Lambda}$Zr. It turns out that although the 
presence of the $\Lambda$ lowers the Fermi level,
the $\Lambda$-N interaction does not have enough 
strength to stabilize the core A$_{c}$=140. But
if a single $\Lambda$ cannot stabilize such a large 
core, maybe two $\Lambda$ particles could. An interesting
question is whether such objects could be found in the 
environment in which the processes of nucleosynthesis 
occur.

It is important to understand the microscopic 
mechanism through which the $\Lambda$ binds the 
additional pair of neutrons in $^{42+\Lambda}$Ne.
In a recent study~\cite{LVR.97} we have used the 
relativistic Hartree-Bogoliubov model to analyze
the isospin dependence 
of the spin-orbit interaction in light neutron-rich
nuclei. It has been shown 
that the magnitude of the spin-orbit 
potential is considerably reduced in drip line nuclei.
With the increase of the neutron number, the effective 
one-body spin-orbit potential becomes weaker. This 
results in a reduction of the energy splittings between 
spin-orbit partners. The reduction of the spin-orbit 
potential is especially pronounced in the surface region.
In the relativistic mean-field
approximation, the spin-orbit potential originates from the addition
of two large fields: the field of the vector mesons (short
range repulsion), and the scalar field of the sigma meson (intermediate
attraction). In the first order approximation, and assuming spherical 
symmetry, the spin orbit term can be written as
\begin{equation}
\label{so1}
V_{s.o.} = {1 \over r} {\partial \over \partial r} V_{ls}(r), 
\end{equation} 
where $V_{ls}$ is the spin-orbit potential~\cite{Rin.96,Koepf.91}
\begin{equation}
\label{so2}
V_{ls} = {m \over m_{eff}} (V-S).
\end{equation}
V and S denote the repulsive vector and the 
attractive scalar potentials, respectively.
$m_{eff}$ is the effective mass
\begin{equation}
\label{so3}
m_{eff} = m - {1 \over 2} (V-S).
\end{equation}

Using the vector and scalar potentials from the self-consistent
ground-state solutions, we have computed from~(\ref{so1}) - (\ref{so3}) the 
spin-orbit terms for several Ne isotopes
and corresponding $\Lambda$-hypernuclei.  
They are displayed in Fig. 2  as function of the radial distance from the
center of the nucleus. The magnitude of the spin-orbit term $V_{s.o.}$ in Ne
nuclei (dashed lines) decreases as we add more neutrons, i.e. more units of 
isospin. The reduction for nuclei close to the neutron drip is $\approx 40\%$ 
in the surface region, as compared to values which correspond to beta stable 
nuclei. For the corresponding $\Lambda$-hypernuclei 
(solid lines) the spin-orbit term displays 
an increase in magnitude of about 10\% (smaller as we 
approach the drip line). This is a rather surprising 
result, as we have seen that the single-particle  
energies and bulk properties seem to be less affected 
by the presence of the $\Lambda$ particle. The effect
is purely relativistic and it appears to be strong 
enough to bind an additional pair of neutrons at the 
drip line. The mean field 
potential, in which the nucleons move, results from
the cancelation of two large meson potentials: the attractive scalar potential
S and the repulsive vector potential V: V+S. 
The spin-orbit potential, on the other hand, arises
from the very strong anti-nucleon
potential V-S. Therefore, while
in the presence of the $\Lambda$ the changes in V and S
cancel out in the mean-field potential, they  are amplified in $V_{ls}$. 
We illustrate this effect on the example 
of $^{30}$Ne and the corresponding hypernucleus  
$^{31}_{\Lambda}$Ne. For the core $^{30}$Ne
the values of the scalar (S) and vector (V) potential in 
the center of the nucleus are -380 MeV and 308 MeV,  
respectively. For $^{31}_{\Lambda}$Ne the corresponding
values are: -412 MeV and 336 MeV. The addition of the 
$\Lambda$ particle changes the value of the mean-field
potential in the center of the nucleus by 4 MeV, but it
changes the anti-nucleon potential by 60 MeV.
This is reflected in the corresponding spin-orbit term
(Fig. 2), which provides more binding for states close
to the Fermi surface. The additional binding stabilizes 
the hypernuclear core.  

In conclusion, we report results of the first 
application of the Relativistic Hartree Bogoliubov
model in coordinate space, with finite range pairing 
interaction, to the description 
of $\Lambda$-hypernuclei with a large neutron excess.
In particular, we have studied the effects of the 
$\Lambda$ hyperon in its ground state on Ne nuclei 
with neutron halo.
Although the inclusion of the $\Lambda$ hyperon does
not produce excessive changes in bulk properties, 
we find that it can shift the neutron drip by 
stabilizing an otherwise unbound core nucleus at
drip line. The microscopic mechanism through which
additional neutrons are bound to the core originates
from the increase in magnitude of the spin-orbit 
term in presence of the $\Lambda$ particle. We find that
the $\Lambda$ in its ground state produces only a 
fractional change in the central mean-field potential.
On the other hand, through a purely relativistic 
effect, it notably changes the spin-orbit
term in the surface region, providing additional
binding for the outermost neutrons. Neutron-rich
$\Lambda$-hypernuclei might have an important role
in the process of nucleosynthesis by neutron capture.
\bigskip

This work has been supported by the
Bundesministerium f\"ur Bildung und Forschung under 
project 06 TM 875. G. A. L. acknowledges support 
from DAAD.
\

\newpage

\centerline{\bf Figure Captions}

\begin{itemize}

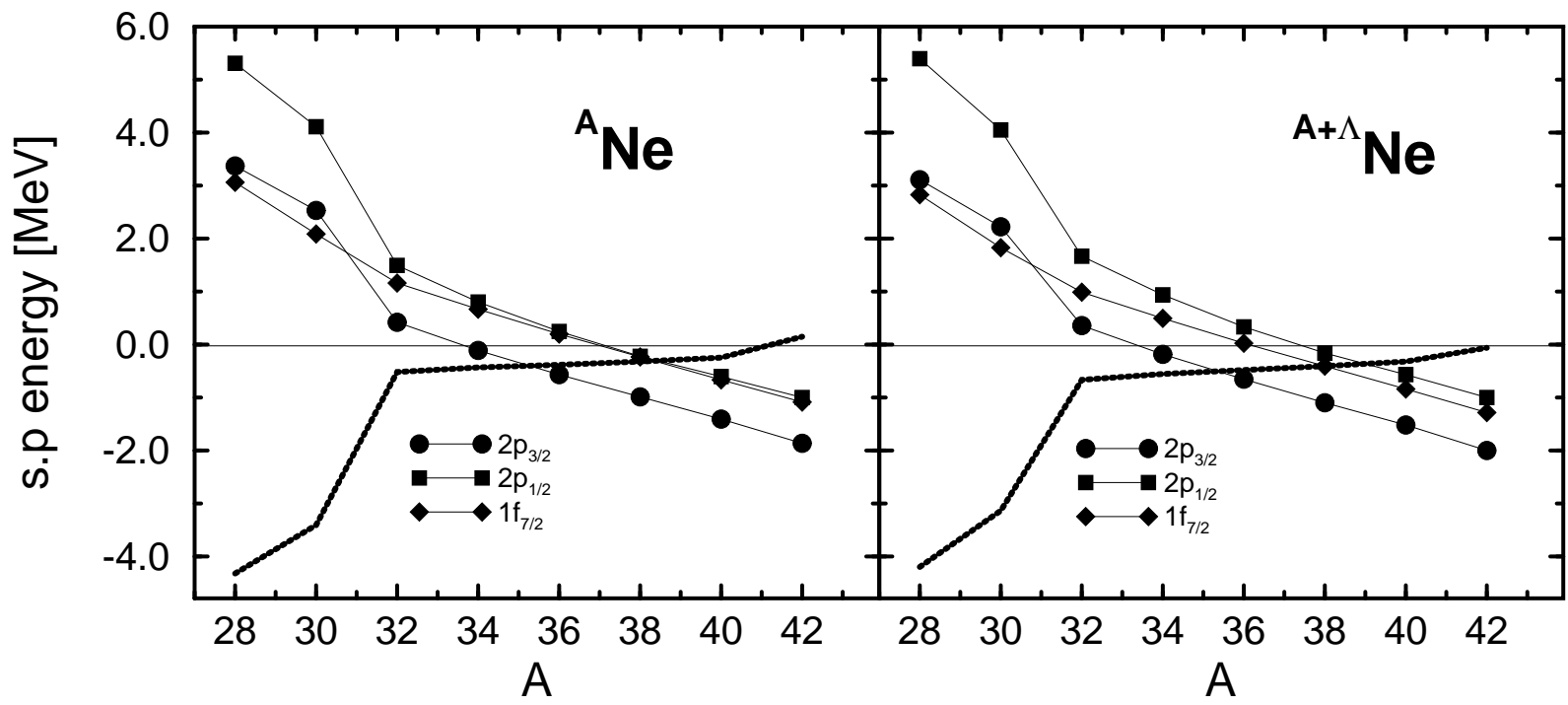
\item{\bf Fig.1} 1f-2p single-particle neutron levels in the canonical 
basis for the $Ne$ (a), and $Ne + \Lambda$ (b) isotopes. The dotted
line denotes the Fermi level.

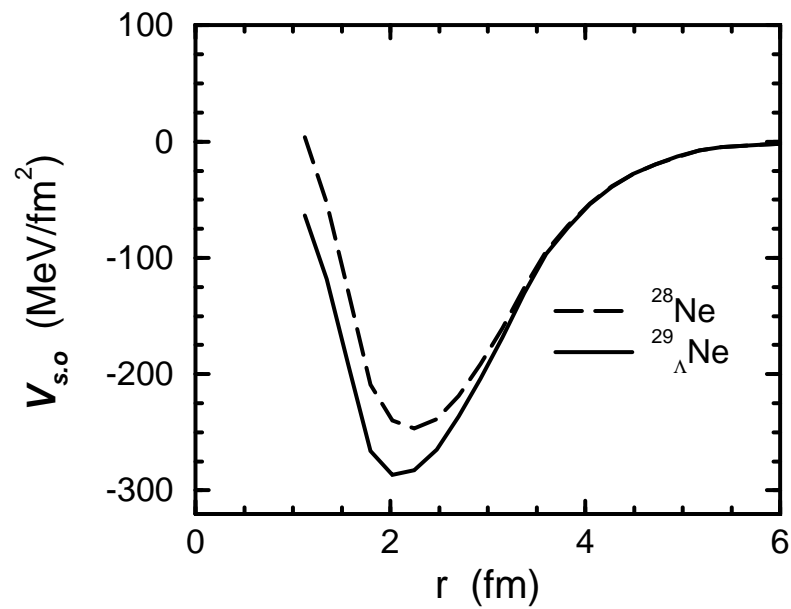
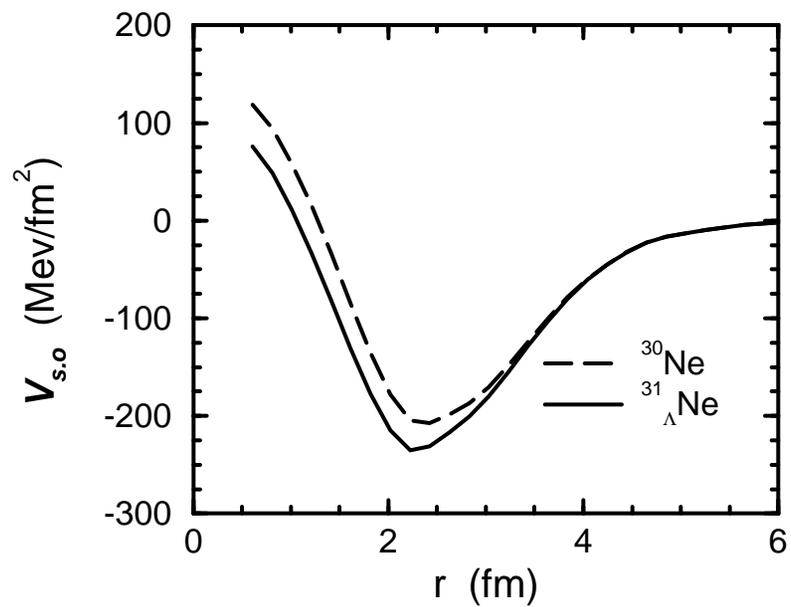
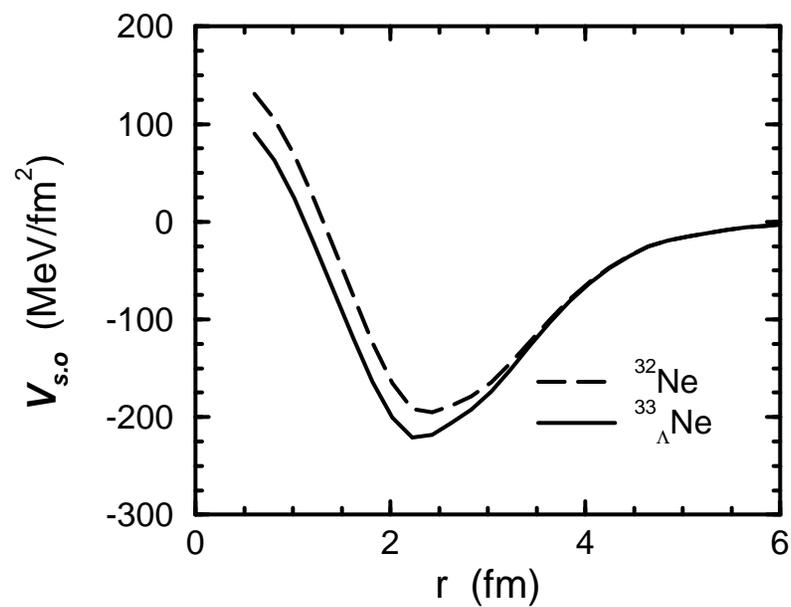
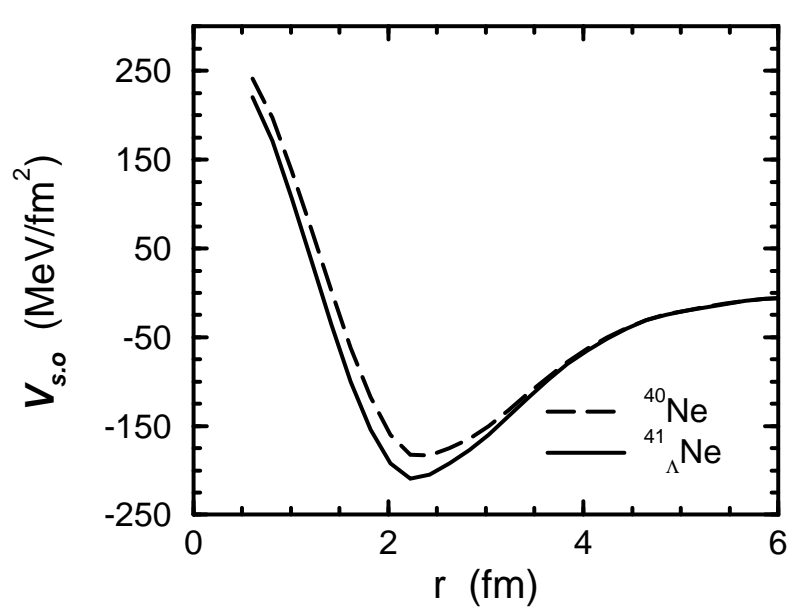
\item{\bf Fig.2} Radial dependence of the spin-orbit potential
in self-consistent solutions for the ground-states of 
$Ne$ (a), and $Ne + \Lambda$ (b) isotopes.

\end{itemize}

\end{document}